# Strain-coupled ferroelectric polarization in $BaTiO_3$-$CaTiO_3$ superlattices

Sung Seok A. Seo and Ho Nyung Lee*

*Materials Science and Technology Division, Oak Ridge National Laboratory, Oak Ridge, TN 37831, USA*

We report on growth and ferroelectric (FE) properties of superlattices (SLs) composed of the FE $BaTiO_3$ and the paraelectric (PE) $CaTiO_3$. Previous theories have predicted that the polarization in $(BaTiO_3)n/(CaTiO_3)n$ SLs increases as the sublayer thickness ($n$) increases when the same strain state is maintained. However, our $BaTiO_3/CaTiO_3$ SLs show a varying lattice-strain state and systematic reduction in polarization with increasing $n$ while coherently-strained SLs with $n$=1, 2 show a FE polarization of ca. 8.5 $\mu C/cm^2$. We suggest that the strain coupling plays more important role in FE properties than the electrostatic interlayer coupling based on constant dielectric permittivities.

* E-mail: hnlee@ornl.gov

Advances in modern oxide synthesis techniques with atomic-level control have shed light on exploring physical properties of artificially-designed ferroelectric (FE) oxide heterostructures and superlattices (SLs).[1-7] For example, enhanced FE and dielectric properties could be achieved in SLs composed of FE $BaTiO_3$ (BTO) and paraelectric (PE) $SrTiO_3$ (STO) layers.[1,2,6,8] Such enhancement of ferroelectric polarization is known to originate from the strong strain coupling of ferroelectric polarization in $BaTiO_3$-based ferroelectrics.[9,10] Recently, by including PE $CaTiO_3$ (CTO) layers, three-component SLs have been designed and raised an intriguing issue of artificially broken compositional inversion symmetry,[11,12] resulting enhanced ferroelectric properties.[7,13] Along with these experimental developments, recent progress in computational methods[14-16] have provided us with opportunities to systematically investigate these FE oxide SLs as well.[17-20] Especially, Nakhmanson *et al.* performed *ab initio* calculations combined with a genetic algorithm technique to optimize the arrangement of individual BTO, STO, and CTO layers in a SL form, and found that (BTO)*n*/(CTO)*n* SLs, where *n* is the number of unit-cells (u.c.) in each BTO and CTO layers, reach the highest polarization for the largest value of *n* when perfect strain is maitained.[20] The (BTO)*n*/(CTO)*n* SLs have the same chemical composition as the $Ba_{1-x}Ca_xTiO_3$ solid-solution at $x$=0.5, but the alloy cannot exist naturally due to the solubility limit.[21] The considerably different ionic sizes of Ba and Ca not only make very different lattice parameters of BTO and CTO but also result in the phase segregation thermodynamically stable. Hence, the growth of the (BTO)*n*/(CTO)*n* SLs might be a formidable task in experiment although it would be interesting to experimentally check the ferroelectric properties in BTO/CTO SLs with various combinations of sublayer thickness. By using a controlled layer-by-layer growth technique, we have previously fabricated a (BTO)1/(CTO)1 SL, which revealed that the polarization was indeed three times larger than that of (BTO)1/(STO)1 SL.[22] We also suggested a theoretical explanation of the increased polarization as a result of the atomic corrugation of CaO layers, which plays an important role in stabilizing the FE state in the (BTO)1/(CTO)1 SL. However, a systematic study with increasing *n* is still required to understand the polarization coupling in (BTO)*n*/(CTO)*n* SLs.

In this letter, we report on the growth and FE properties of (BTO)*n*/(CTO)*m* SLs, with *n,m*=1−5 u.c. Atomic-scale (BTO)*n*/(CTO)*m* SLs were synthesized on STO (001) single-crystalline substrates using pulsed laser deposition equipped with reflection high-energy electron diffraction. For electrical characterization, epitaxial $SrRuO_3$ conducting thin films were deposited as bottom electrodes with preserved single terrace steps (~0.4 nm) on STO substrates,[23] which also provided an essential base for growing SLs with nearly-abrupt interfaces. Since the thickness of the SRO layer was typically less than 500 Å, its in-plane lattice constant was found to be the same as that of STO, *i.e.* 3.905 Å. Then, we grew BTO and CTO layers up to about 200 nm in total thickness. Note that single-crystalline BTO is tetragonal ($a$=$b$=3.994 Å, $c$=4.033 Å)[24] and single-crystalline CTO is orthorhombic ($a$=5.442 Å, $b$=5.380 Å, $c$=7.640 Å, and pseudo-cubic $a_c$=3.826 Å)[25] at room temperature. The alternation of compressive and tensile strains, which are induced in BTO and CTO layers by STO (cubic, $a$=3.905 Å) substrates, respectively, can effectively cancel out their opposite-directional mechanical tension, so the (BTO)$n$/(CTO)$n$ SLs might be expected to better preserve a fully strained state when the layer thicknesses are properly combined, as shown in Fig. 1 (a). This kind of oxide heterostructure can be as an ideal prototype for investigating FE and PE coupling in layered structures.

Figure 1(b) shows x-ray diffraction $\theta$-$2\theta$ scans of a (BTO)5/(CTO)5 SL. The asterisks originate from the 001 and 002 Bragg peaks of the STO substrate. The well-defined 00$l$ peaks are presented due to the artificial periodicity of the SL ($l$: integers). The full-width half-maxima in $\omega$ scans of SL peaks are all less than 0.05°, confirming the high crystallinity of our SLs. These clear satellite peaks demonstrate that the SL structures are grown well with the artificial periodicities along the [001]-direction with well-defined interfaces.

To investigate FE properties of the (BTO)$n$/(CTO)$m$ SLs, we recorded polarization *vs.* electric field ($P$–$E$) hysteresis loops using a TF analyser (aixACCT) at room temperature. Figure 2(a) shows typical FE $P$–$E$ hysteresis loops with different combinations of $n$ and $m$. In order to make our discussion concise, here we focus on the behavior of polarization while leaving the coercive field change aside. We observe the highest values of the remanent polarization ($P_r$), $P_r \equiv (+P_r+(-P_r))/2$, ~8.5 μC/cm$^2$ for ($n$, $m$)=(1, 1) and (2, 2) among our samples. As the sublayer thickness of (BTO)$n$/(CTO)$m$ SLs increases, $P_r$ decreases, for example, $P_r$ of (BTO)4/(CTO)4 reaches ~3.1 μC/cm$^2$.

From an experimental point of view, this observation that the one- and two-unit-cell-layer-thick SLs have the highest value of $P_r$ is very interesting because growth of such short-period SLs should require the most delicate control of over 100 interfaces. Moreover, shorter period SLs are theoretically expected to result in a reduced polarization due to increased hardening and modification of local soft modes at the interface.[26] It is noteworthy that the highest $P_r$ value is about four times larger than that of pseudo-binary alloy of (*0.5*)BTO–(*0.5*)CTO, *i.e.* $P_r \approx 2$ μC/cm$^2$ (Ref. 27). It suggests that the intermixing of Ba and Ca ions is negligibly small at the interfaces of our SLs even for the (BTO)1/(CTO)1 SL as is also confirmed by the presence of well-defined SL peaks in x-ray scans.

Figure 2(b) shows the change in $P_r$, which is normalized by $P_r$ of relaxed BTO thin films (~11.1 μC/cm$^2$, Ref. 7), as a function of the approximated thickness ratio ($\beta$) between BTO and CTO sublayers, i.e. $\beta \equiv n/m \approx t_{BTO}/t_{CTO}$ for the (BTO)*n*/(CTO)*m* SLs. Our data looks dispersed without any systematic trend despite the fact that $P_r$ of (BTO)*n*/(CTO)*m* SLs is expected to increase gradually as BTO sublayer thickness increases as predicted by first-principles calculations[20] and experimentally observed in other SLs.[7] Figure 2(b) shows the calculated polarization values taken from Ref. 20, which are also normalized by the calculated polarization value of BTO. The calculated values are overall higher than our experimental data. This might be due to the fact that the polarization of BTO used for the calculation is higher than the experimentally recorded one and the calculation excludes the possible change in strain state. We also note that an epitaxial CTO film grown on STO under tensile strain remains paraelectric along the out-of-plane direction at room temperature as we do not observe a notable ferroelectric *P-E* loop in our experiment (data not shown).

In order to take into account the paraelectricity in CTO, an electrostatic model[17] is useful to see the $\beta$-dependence of $P_r$ for FE/PE SL structures, as follows:

$$P_r = P_{FE}/(1+(\varepsilon_{(FE)}/\varepsilon_{(PE)})/\beta). \qquad (1)$$

Here, $P_{FE}$ is the polarization of FE layer (BTO), and $\varepsilon_{(FE)}/\varepsilon_{(PE)}$ is the dielectric permittivity ratio between the FE and PE layers. The dielectric permittivity values are reported to be ~150 for both CTO thin films[28] and BTO thin films[29] on STO substrates at room temperature, yielding $\varepsilon_{(FE)}/\varepsilon_{(PE)} \approx 1$.

Figure 2(b) also shows the $\beta$-dependence on normalized $P_r$, clearly showing that our experimental data are rather inconsistent with the simple electrostatic model.

However, the large discrepancy between experimentally observed $P_r$ values and calculated ones can be attributed to the difference between experimental and theoretical circumstances. For example, our SLs' lattice constants and strain states are not the same as used in theoretical calculations, in which they are assumed to be coherently strained to STO substrate. As is already known in $BaTiO_3$-based SLs, the ferroelectric polarization couples to strain rather strongly.[7,9,30]

Therefore, we performed x-ray reciprocal space mappings near the off-specular 114-reflection in order to check the strain state as displayed in Fig. 3. The in-plane lattice constants of the (BTO)1/(CTO)1 and (BTO)2/(CTO)2 SLs are closely matched with those of the STO substrates, confirming that the in-plane lattices of these short-periodicity SLs are fully strained. On the other hand, SL samples with longer periodicities show the in-plane strain relaxation, which is defined as the parallel lattice mismatch $(a_{SL}-a_{STO})/a_{STO}\times 100$ (%), where $a_{SL}$ and $a_{STO}$ are the in-plane lattice constants of SL and STO substrate, respectively. For example, the lattice relaxation of (BTO)5/(CTO)5 SL is around +0.46% as shown in Fig. 3(b). Since the strain relaxation is still smaller than the lattice mismatch +2.3% between bulk BTO and STO, it suggests that the in-plane lattices become only partially relaxed, introducing misfit dislocations. It is remarkable that the overall in-plane lattice strain relaxation in our SLs occurs only along the positive direction, *i.e.* the direction of the bulk BTO lattice, which means that the strain relaxation is dominated by BTO layers rather than CTO layers. This might be related to the higher crystallographic symmetry of BTO (tetragonal) than that of CTO (orthorhombic) at room temperature, which makes the structural distortion by strain in tetragonal BTO is less tolerant than in orthorhombic CTO.

Figure 4 shows $P_r$ of (BTO)$n$/(CTO)$n$ SLs normalized by $P_r$ of the (BTO)1/(CTO)1 SL as a function of in-plane strain relaxation. In order to avoid confusion between the different $\beta$, we only consider symmetric SLs, *i.e.* SLs with the same sublayer number of unit-cells between BTO and CTO. While the simple electrostatic model predicts a constant $P_r$ value for these symmetric SL structures and the genetic algorithm technique based on first-principles calculation[20] as well as first-principles based effective

Hamiltonian calculation[26] predict an increase in $P_r$ with $n$ when the SLs' physical constants such as dielectric constant and strain are unchanged, our experimental $P_r$ values gradually decrease as the greater strain relaxation evolves with larger $n$. It seems that there is a big disagreement between the experiment and theoretical calculations. However, this discrepancy can be understood by taking into account of the strain effect on the dielectric permittivity[31] and the polarization[7,9]. Moreover, it is reasonable in the sense that a phenomenological theory predicted that, with only 0.4% strain-relaxation, one can increase the dielectric permittivity of BTO ($\varepsilon_{(BTO)}$) about a factor of about 5.[32] Therefore, if we take the dielectric permittivity change into account for the electrostatic model (Eq. (1)), a dramatic decrease of $P_r$ from 8.5 $\mu C/cm^2$ ($\varepsilon_{(BTO)}/\varepsilon_{(CTO)}=1$) to 1.4 $\mu C/cm^2$ ($\varepsilon_{(BTO)}/\varepsilon_{(CTO)}=5$) could be obtained, which now describes qualitatively the experimentally measured $P_r$ values from our SLs. In addition to the dielectric permittivity change and strain relaxation, one can also consider the possible crystallographic symmetry lowering that is also known to greatly influence the ferroelectric properties of ferroelectric-paraelectric SLs.[33,34] However, we could not confirm such symmetry change in our SLs within the resolution of our in-house x-ray diffraction. It is also worthy to note that a recent first-principles study (Ref. 35) suggests possible stabilization of a ferroelectric phase when strong tensile strains are applied to $CaTiO_3$, resulting in a large in-plane polarization. Such a strong development of the in-plane polarization could be responsible for the systematic reduction of out-of-plane polarization observed experimentally in this work.

In conclusion, short-period BTO/CTO SLs were grown on atomically-flat $SrRuO_3$-covered STO substrates by PLD. While first-principles calculations suggested a specific combination of materials and thicknesses for SLs with a maximum polarization, the trend of increase in polarization with increasing thickness was not found. We note a marked difference, which illustrates the importance of strain and its relaxation in larger structures. In fact, strain relaxation here is unique because the average in-plane lattice parameter of the structures does not tend towards the average between the SL's constituents. This shows that the relaxation behavior of BTO is different from that of CTO. Moreover, it is remarkable that the less relaxed shorter-period SLs can stabilize the FE state even better than the longer-period SLs. Hence, both well-strained lattice and proper choice of sublayer thickness in FE/PE heterostructures are essential for enhancing their ferroelectric properties.

We thank V. R. Cooper, H. M. Christen, and K. M. Rabe for useful discussions and comments. This work was sponsored by the Division of Materials Sciences and Engineering, U. S. Department of Energy.

# References

[1] K. Iijima, T. Terashima, Y. Bando, K. Kamigaki, and H. Terauchi, J. Appl. Phys. **72**, 2840 (1992).

[2] H. Tabata, H. Tanaka, and T. Kawai, Appl. Phys. Lett. **65**, 1970 (1994).

[3] H. M. Christen, L. A. Boatner, J. D. Budai, M. F. Chisholm, L. A. Gea, P. J. Marrero, and D. P. Norton, Appl. Phys. Lett. **68**, 1488 (1996).

[4] J. C. Jiang, X. Q. Pan, W. Tian, C. D. Theis, and D. G. Schlom, Appl. Phys. Lett. **74**, 2851 (1999).

[5] D. G. Schlom, J. H. Haeni, J. Lettieri, C. D. Theis, W. Tian, J. C. Jiang, and X. Q. Pan, Mater. Sci. Eng., B **87**, 282 (2001).

[6] M. Dawber, K. M. Rabe, and J. F. Scott, Rev. Mod. Phys. **77**, 1083 (2005).

[7] H. N. Lee, H. M. Christen, M. F. Chisholm, C. M. Rouleau, and D. H. Lowndes, Nature **433**, 395 (2005).

[8] D. A. Tenne, A. Bruchhausen, N. D. Lanzillotti-Kimura, A. Fainstein, R. S. Katiyar, A. Cantarero, A. Soukiassian, V. Vaithyanathan, J. H. Haeni, W. Tian, D. G. Schlom, K. J. Choi, D. M. Kim, C. B. Eom, H. P. Sun, X. Q. Pan, Y. L. Li, L. Q. Chen, Q. X. Jia, S. M. Nakhmanson, K. M. Rabe, and X. X. Xi, Science **313**, 1614 (2006).

[9] K. J. Choi, M. Biegalski, Y. L. Li, A. Sharan, J. Schubert, R. Uecker, P. Reiche, Y. B. Chen, X. Q. Pan, V. Gopalan, L. Q. Chen, D. G. Schlom, and C. B. Eom, Science **306**, 1005 (2004).

[10] D. G. Schlom, L.-Q. Chen, C.-B. Eom, K. M. Rabe, S. K. Streiffer, and J.-M. Triscone, Annu. Rev. Mater. Sci. **37**, 589 (2007).

[11] N. Sai, B. Meyer, and D. Vanderbilt, Phys. Rev. Lett. **84**, 5636 (2000).

[12] M. P. Warusawithana, E. V. Colla, J. N. Eckstein, and M. B. Weissman, Phys. Rev. Lett. **90**, 036802 (2003).

[13] D. A. Tenne, H. N. Lee, R. S. Katiyar, and X. X. Xi, J. Appl. Phys. **105**, 054106 (2009).

[14] R. D. Kingsmith and D. Vanderbilt, Phys. Rev. B **47**, 1651 (1993).

[15] O. Diéguez, K. M. Rabe, and D. Vanderbilt, Phys. Rev. B **72**, 144101 (2005).

[16] X. Wu, M. Stengel, K. M. Rabe, and D. Vanderbilt, Phys. Rev. Lett. **101**, 087601 (2008).

[17] J. B. Neaton and K. M. Rabe, Appl. Phys. Lett. **82**, 1586 (2003).

[18] L. Kim, J. Kim, D. Jung, J. Lee, and U. V. Waghmare, Appl. Phys. Lett. **87**, 052903 (2005).

[19] S. M. Nakhmanson, K. M. Rabe, and D. Vanderbilt, Appl. Phys. Lett. **87**, 102906 (2005).

[20] S. M. Nakhmanson, K. M. Rabe, and D. Vanderbilt, Phys. Rev. B **73**, 060101 (2006).

[21] T. Mitsui and W. B. Westphal, Phys. Rev. **124**, 1354 (1961).

[22] S. S. A. Seo, J. H. Lee, H. N. Lee, M. F. Chisholm, W. S. Choi, D. J. Kim, J. Y. Jo, H. Kim, J. Yu, and T. W. Noh, Adv. Mater. **19**, 2460 (2007).

[23] H. N. Lee, H. M. Christen, M. F. Chisholm, C. M. Rouleau, and D. H. Lowndes, Appl. Phys. Lett. **84**, 4107 (2004).
[24] H. D. Megaw, Acta Cryst. **15**, 972 (1961).
[25] S. Sasaki, C. T. Prewitt, J. D. Bass, and W. A. Schulze, Acta Cryst. C **43**, 1668 (1987).
[26] J. H. Lee, J. Yu, and U. V. Waghmare, J. Appl. Phys. **105**, 016104 (2009).
[27] X. S. Wang, H. Yamada, and C. N. Xu, Appl. Phys. Lett. **86**, 022905 (2005).
[28] J. Hao, W. Si, X. X. Xi, R. Guo, A. S. Bhalla, and L. E. Cross, Appl. Phys. Lett. **76**, 3100 (2000).
[29] T. Shimuta, O. Nakagawara, T. Makino, S. Arai, H. Tabata, and T. Kawai, J. Appl. Phys. **91**, 2290 (2002).
[30] H. N. Lee, S. M. Nakhmanson, M. F. Chisholm, H. M. Christen, K. M. Rabe, and D. Vanderbilt, Phys. Rev. Lett. **98**, 217602 (2007).
[31] J. H. Haeni, P. Irvin, W. Chang, R. Uecker, P. Reiche, Y. L. Li, S. Choudhury, W. Tian, M. E. Hawley, B. Craigo, A. K. Tagantsev, X. Q. Pan, S. K. Streiffer, L. Q. Chen, S. W. Kirchoefer, J. Levy, and D. G. Schlom, Nature **430**, 758 (2004).
[32] N. A. Pertsev, A. G. Zembilgotov, and A. K. Tagantsev, Phys. Rev. Lett. **80**, 1988 (1998).
[33] S. Rios, A. Ruediger, A. Q. Jiang, J. F. Scott, H. Lu, and Z. Chen, J. Phys.: Condens. Matter **15**, L305 (2003).
[34] K. Johnston, X. Huang, J. B. Neaton, and K. M. Rabe, Phys. Rev. B **71**, 100103 (2005).
[35] C.-J. Eklund, C. J. Fennie, and K. M. Rabe, cond-mat.mtrl-sci, arXiv:0904.2118v (2009).

## Figures & captions

**Figure 1.** (a) Lattice mismatches of $CaTiO_3$ and $BaTiO_3$ on $SrTiO_3$ by considering their pseudo-cubic lattice constants. Schematic diagram of a $BaTiO_3/CaTiO_3$ superlattice on a $SrRuO_3/SrTiO_3$ substrate without strain relaxation. Green and red arrows indicate directions of in-plane, biaxial compressive and tensile strains induced in $BaTiO_3$ and $CaTiO_3$ layers, respectively. (b) X-ray $\theta$-$2\theta$ diffraction pattern of a $(BaTiO_3)5/(CaTiO_3)5$ SL. Well defined SL peaks due to artificial periodicity is noted by the 00*l*-reflections. The asterisks (*) indicate peaks from the $SrTiO_3$ substrate.

**Figure 2.** (a) *P-E* curves recorded from $BaTiO_3/CaTiO_3$ superlattices, measured at room temperature. (b) Remanent polarization ($P_r$) as a function of the number of unit-cell ratio ($\beta$) between $BaTiO_3$ and $CaTiO_3$ layers, i.e. $\beta \equiv n/m \approx t_{BTO}/t_{CTO}$. Circles are theoretically calculated values from Ref. 20. The dashed curve represents the calculated $P_r$ by using the electrostatic model (Eq. (1)) as a function of $\beta$.

**Figure 3.** X-ray reciprocal space maps around the 114-reflection of $SrTiO_3$. While the in-plane lattices of a $(BaTiO_3)2/(CaTiO_3)2$ superlattice are fully strained with respect to the $SrTiO_3$ substrates (sold lines), a lattice-strain relaxation of 0.46% is observed from a $(BaTiO_3)5/(CaTiO_3)5$ superlattice (dashed line). The artificially modulated periodicity along the [001] direction and high crystallinity that can be confirmed by the well-defined peaks from Cu-$K\alpha_1$ and $K\alpha_2$ radiations are clearly represented in the reciprocal space maps.

**Figure 4.** Change in $P_r$ as a function of in-plane strain relaxation for symmetric $(BaTiO_3)n/(CaTiO_3)n$ superlattices. The circles are theoretical values from Ref. 20 calculated for ideal $(BaTiO_3)n/(CaTiO_3)n$ superlattices without the in-plane strain relaxation, and the dashed curve represents a fixed polarization value from the electrostatic model.

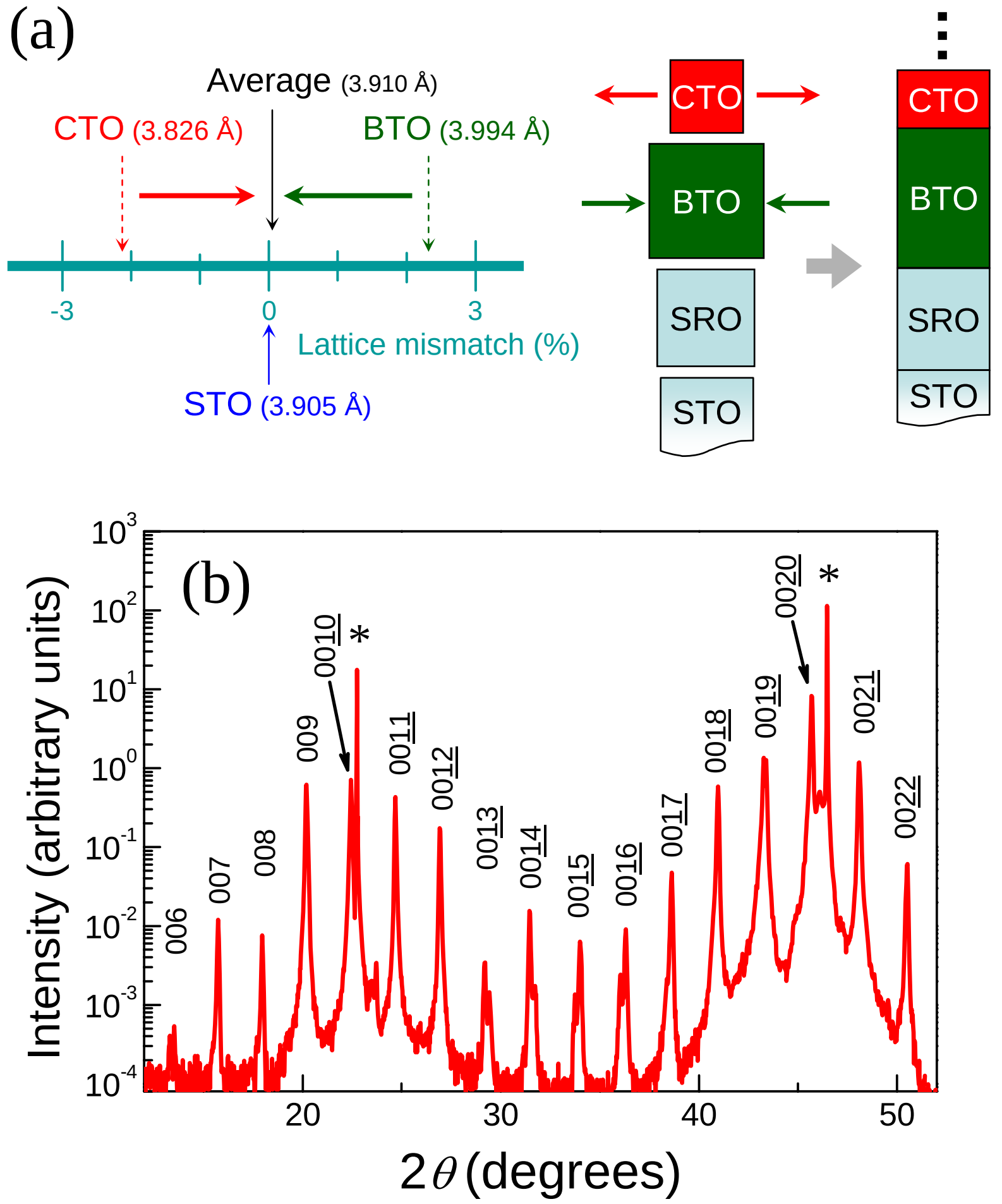

Figure 1.

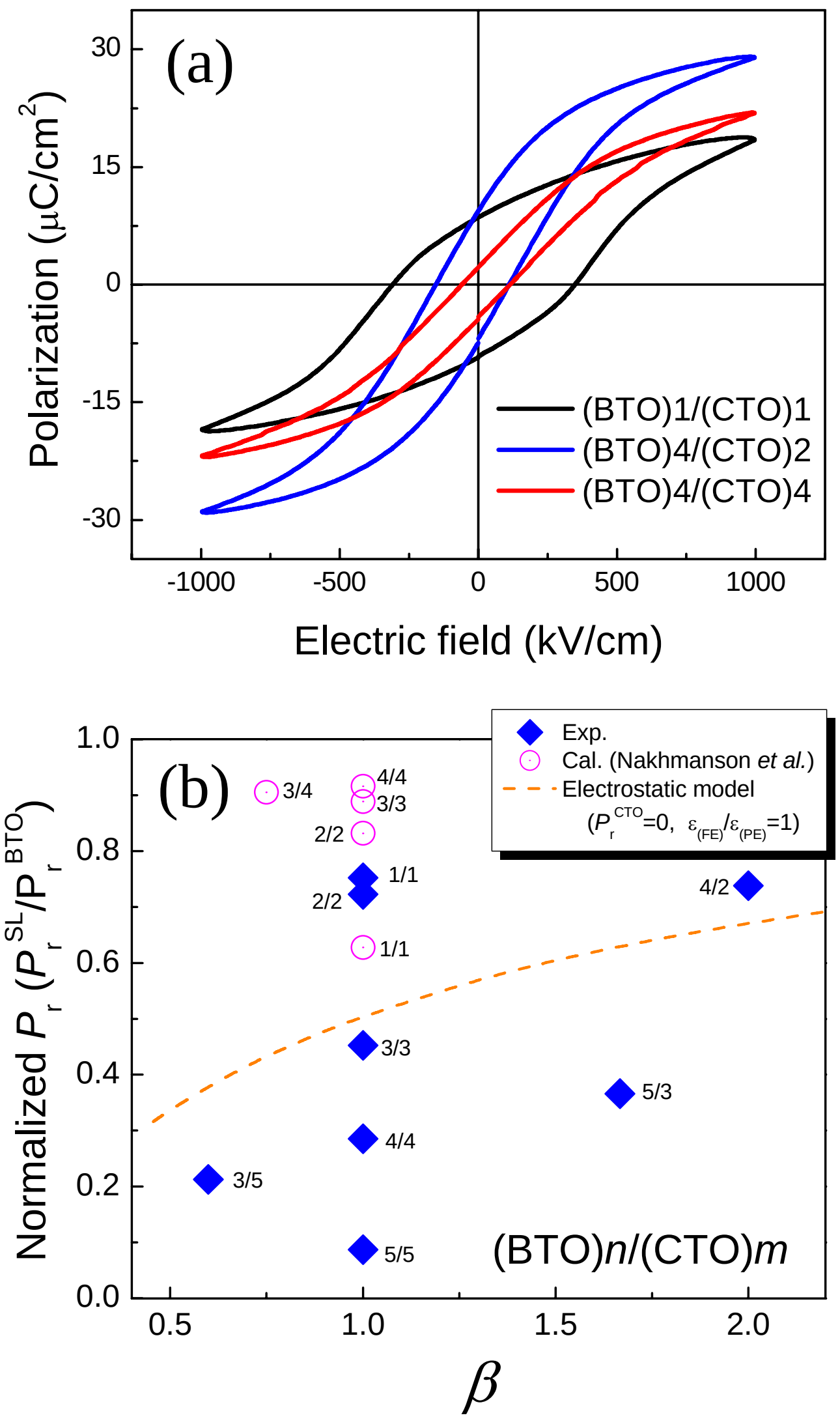

Figure 2.

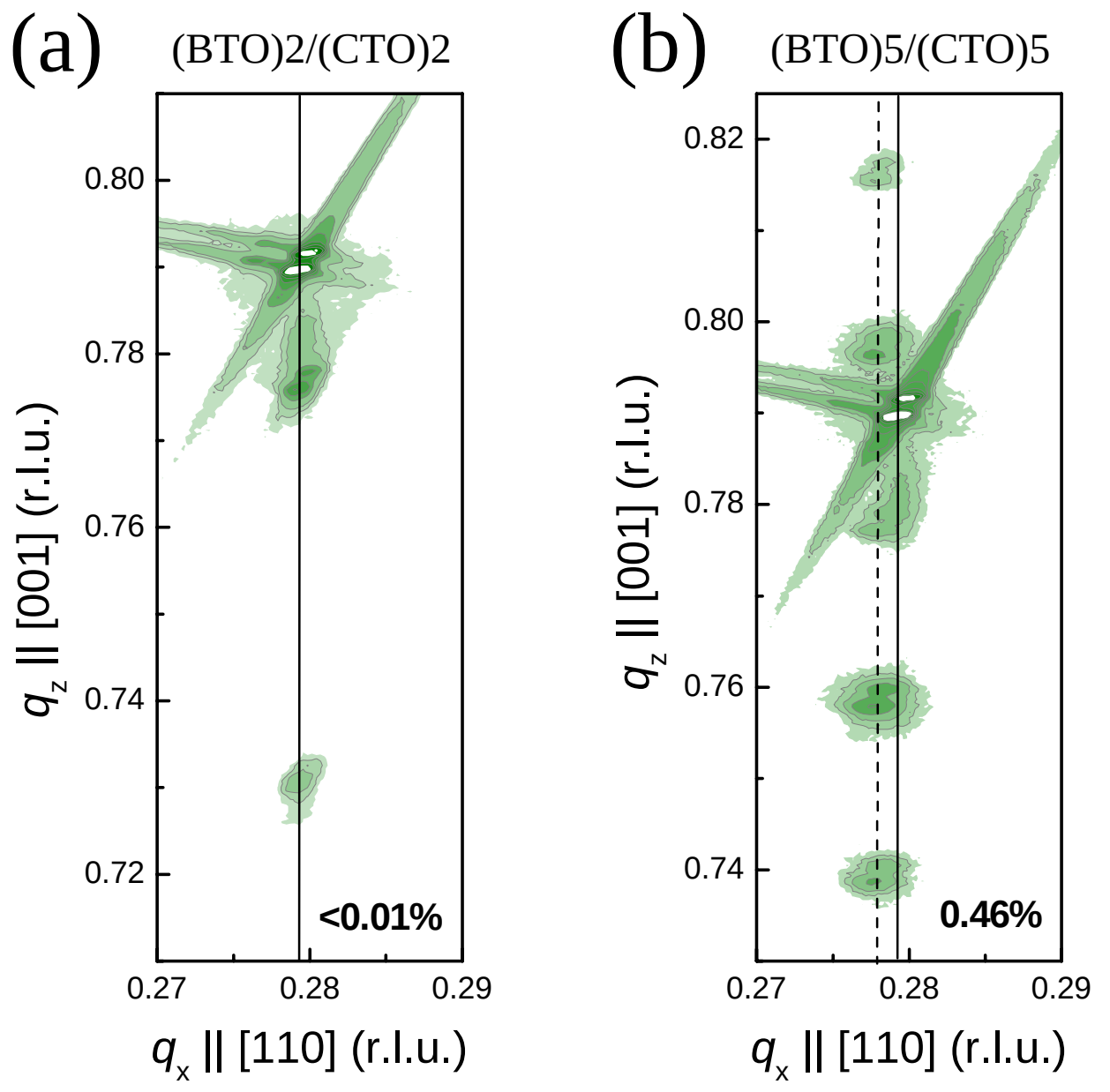

Figure 3.

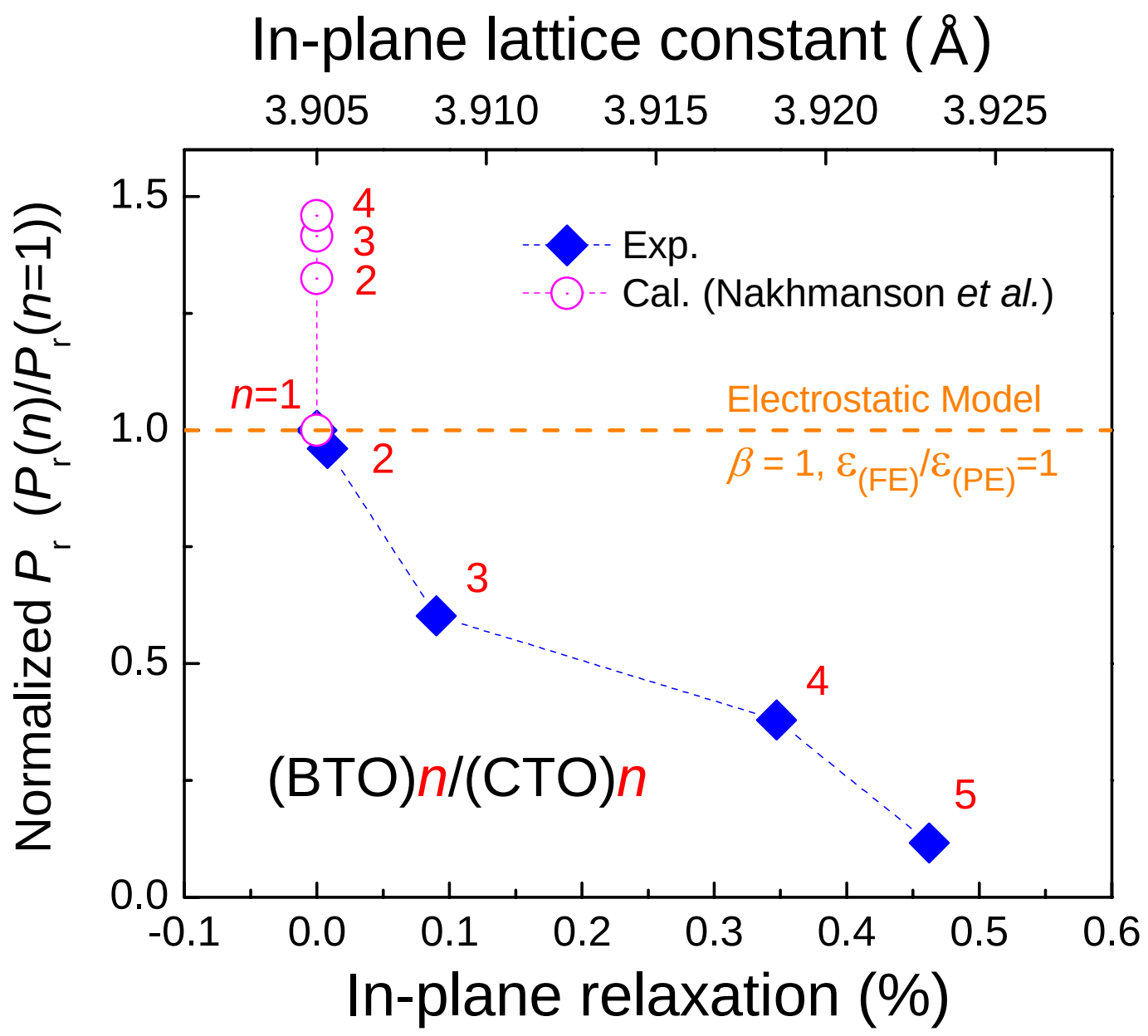

Figure 4.